# Comparison of BERT vs GPT

*Although GPT receives all the publicity, is BERT the new XGBOOST?*

Edward Sharkey, Philip Treleaven
Computer Science
University College London
Edward.sharkey@ucl.ac.uk

## ABSTRACT

The paper benchmarks several Transformer models [4], to show how these models can judge sentiment from a news event. This signal can then be used for downstream modelling and signal identification for commodity trading. We find that fine-tuned BERT models outperform fine-tuned or vanilla GPT models on this task. Transformer models have revolutionized the field of natural language processing (NLP) in recent years, achieving state-of-the-art results on various tasks such as machine translation, text summarization, question answering, and natural language generation. Among the most prominent transformer models are Bidirectional Encoder Representations from Transformers (BERT) and Generative Pre-trained Transformer (GPT), which differ in their architectures and objectives.

This paper introduces the BERT and GPT historical architectures as building blocks to the paradigm shifting transformer models. We then benchmark four models for commodity forecasting: FinBERT [10], CopBERT used for copper , GPT [19] and CopGPT . We compare them to a fine-tuned commodity model CopBERT. To disambiguate our XGboost[2] metaphor. XGboost is used to represent a combination of accuracy and interpretability. XGboost is well known as the pinnacle of accuracy for tabular data problems, combined with excellent interpretability through salient feature functionality.

A CopBERT model training data and process overview is provided. The CopBERT model outperforms similar domain specific BERT trained models such as FinBERT. The below confusion matrices show the performance on CopBERT & CopGPT respectively. We see a ~10 percent increase in f1_score when compare CopBERT vs GPT4 and 16 percent increase vs CopGPT. Whilst GPT4 is dominant It highlights the importance of considering alternatives to GPT models for financial engineering tasks, given risks of hallucinations, and challenges with interpretability.

We unsurprisingly see the larger LLMs outperform the BERT models, with predictive power. In summary BERT is partially the new XGboost, what it lacks in predictive power it provides with higher levels of interpretability. Concluding that BERT models might not be the next XGboost [2], but represent an interesting alternative for financial engineering tasks, that require a blend of interpretability and accuracy.

## 1 Introduction

This paper benchmarks domain specific BERT and GPT models for commodities:

- **Bidirectional Encoder Representations from Transformers (BERT)** - is based on Transformers, a deep learning model designed to pretrain deep bidirectional representations [6].
- **Copper BERT (CopBERT)** – Is a retuned BERT model focussed on copper news items for the NLP sentiment analysis task.
- **Copper GPT (CopGPT) -** Is a fine tuned with few short learning [19] GPT model focussed on copper news items for the NLP sentiment analysis task.

In order to help contextualise the finetuned models in this paper, we also introduce the Transformer Tree diagram (see Figure 1), which shows how well-known large language models are split across encoder, decoder, and encoder decoder architecture.

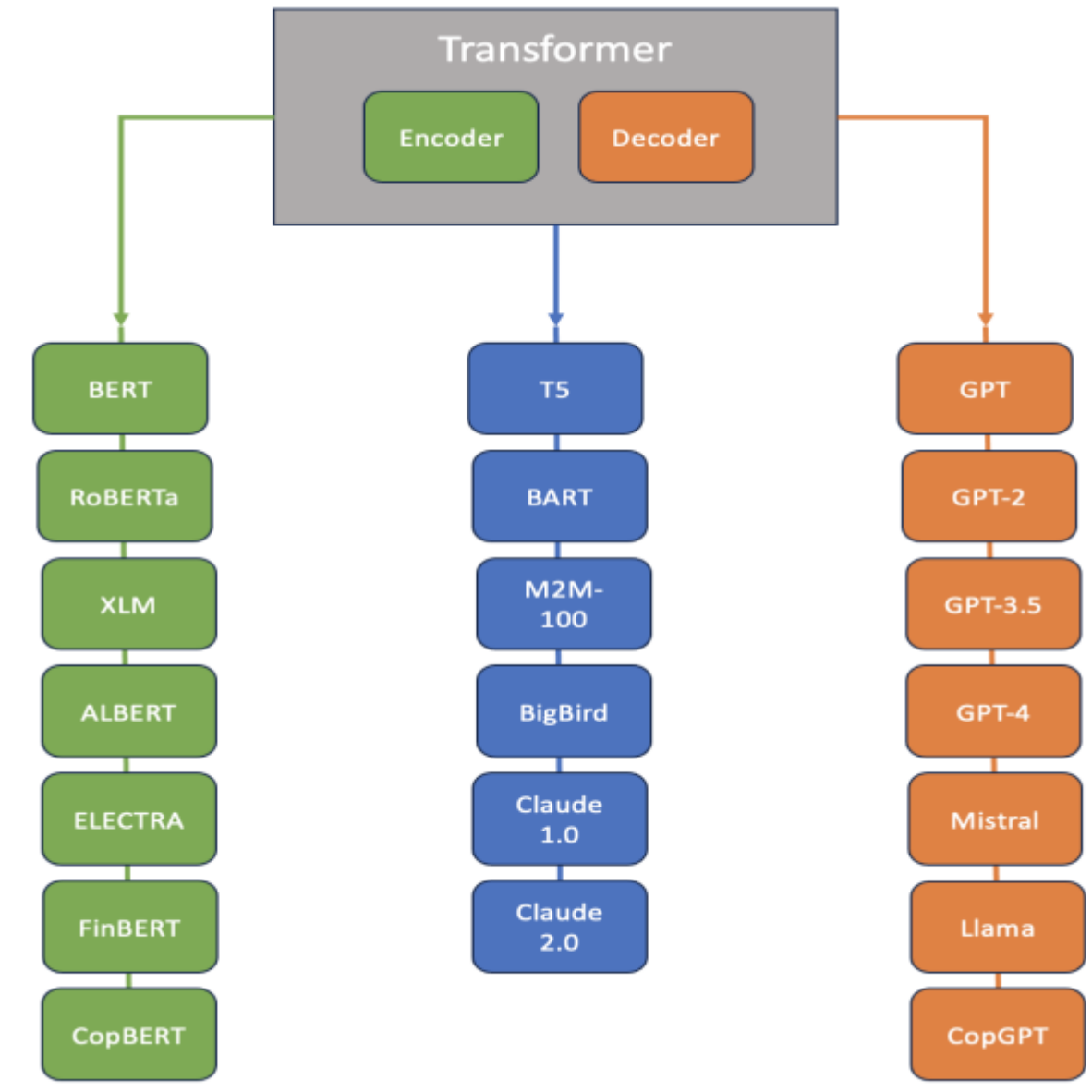

**Figure 1 Transformer Tree Diagram showing popular LLMs by encoder ,decoder and encoder & decoder architecture**

Like other domain specific BERT & GPT models [1], [9], [2], [18] the paper specifically focuses on finetuning a GPT 3.5 model with some prompt engineering combined with a finetuned BERT model [6] developed by Google Brain and GPT [18] developed by OpenAI. To harness transformers for the NLP task of sentiment modelling news article headlines. In this paper, we compare and contrast the respective encoder, and decoder-only architectures of BERT and GPT, and discuss their benefits and limitations for specific use cases. The paper seeks to investigate specific BERT models to assess viability in using commodity market news to understand sentiment, that might impact commodity pricing. The paper highlights the importance of mechanistic interpretability and creates a graph network based on cosine similarity to compare attention maps for CopBERT & CopGPT. The motivation for the research is to address the issues of companies struggling to make correct hedging decisions for the commodities market, in the recent volatile global market. The paper acknowledges the challenges of forecasting models and widely reported challenges with handling the recent "Black Swan" events.

## 2 Transformer Preliminary

As background, the paper presents the fundamental underlying model architecture and introduces self-attention mechanisms, discussing their importance for managing state models. It also highlights the potential benefits of large language models [1] [16] to serve alternative data sets to time series models as part of the augmentation of predictive models. As background:

- **Recurrent Neural Network (RNN) -** outputs of RNN y(tx) are functions of the inputs from previous time steps. It can be expressed or explain as memory functionality or the ability to preserve a level state across time steps.
- **Long Short-Term Memory (LSTM) -** LSTMs are models that are built on RNN models to address short term memory issues. The model can retain and then forget sequence elements.
- **Transformer –** a ground-breaking neural network model that learns context by which is based solely on attention mechanisms, dispensing with recurrency.
- Bidirectional Encoder Representations from **Transformers (BERT) –** as discussed, is based on Transformers, a deep learning model designed to pretrain deep bidirectional representations.
- **Generative Pretrained Transformer -** GPT is primarily used for generating coherent and contextually relevant text. It excels in tasks such as language generation, text completion, and story writing.

## 3 Models

As discussed in the seminal paper "Attention is all you need" [4] Vashani et al discuss the dominant sequence transduction models based on complex recurrent or convolutional neural networks that include an encoder and a decoder. The best performing models also connect the encoder and decoder through an attention mechanism. They propose a new simple network architecture, the Transformer which is based solely on attention mechanisms, dispensing with recurrency. As highlighted in the initial "Attention is all you need" paper [4] the encoder maps an input sequence of symbol representations $(x1,...,xn)$ to a sequence of continuous representations $z = (z1,...,zn)$. Given z, the decoder then generates an output sequence $(y1,...,ym)$ of symbols one element at a time. At each step the model is auto-regressive, consuming the previously generated symbols as additional input when generating the next. The Transformer follows this overall architecture using stacked self-attention and pointwise, fully connected layers for both the encoder and decoder.

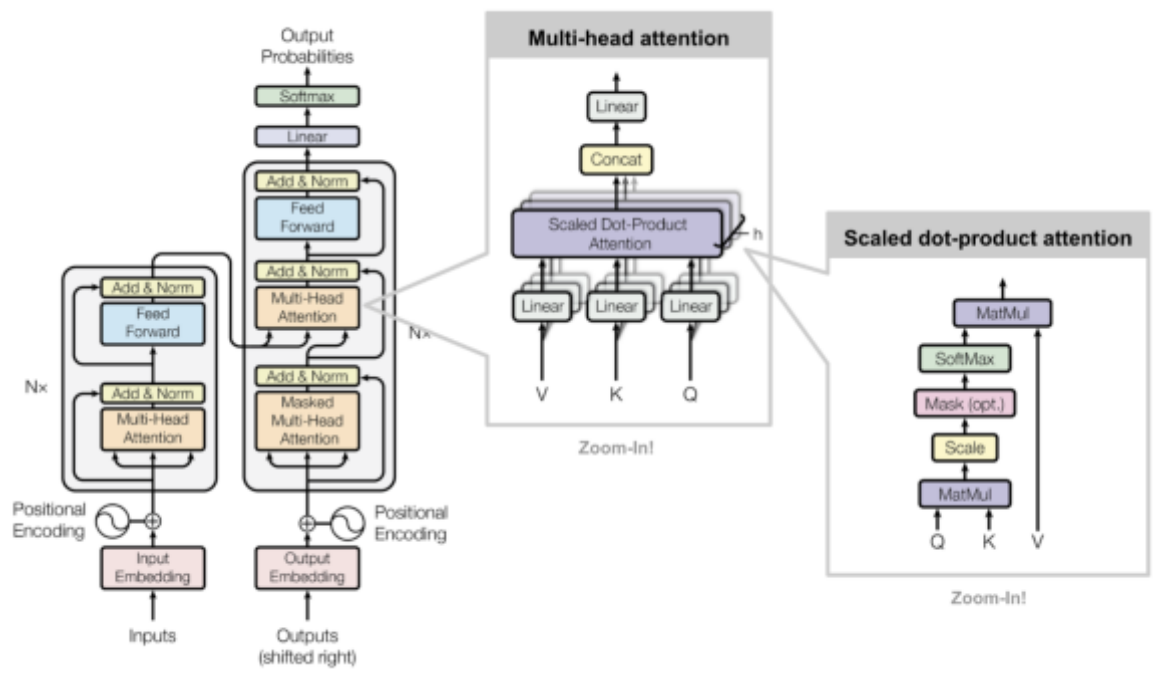

**Figure 2 Transfomer Model with multi-head attention and scaled dot product attention**

Where $Q$ represents a matrix of queries, $K$ represents a matrix of key, $V$ represents a matrix of values. and $d_k$ is the denotes the dimensionality of the queries and keys. The softmax function ensures the output values are between 0 and 1, this what is interpreted as “attention scores”.

As discussed, BERT (Bidirectional Encoder Representations from Transformers) [6] and GPT (Generative Pretrained Transformer) [7] are both transformer-based models developed by Google and OpenAI, respectively. They share a common foundation in the Transformer model introduced by [4] but differ significantly in their structure and training objectives.

We however argue general-purpose models are not effective enough because of the specialised language used in a financial context. FinBERT [8] and BloombergGPT[9] is an example of how this is addressed. FinBERT is a language model based on BERT, to tackle NLP tasks in the financial domain. The results show improvement in every measured metric on current state-of-the-art results for two financial sentiment analysis datasets. FinBERT outperforms state-of-the-art machine learning methods. Hence this paper introduces the concept of CopBERT which seeks to facilitate the NLP task of sentiment analysis to analyse commodity news feeds.

BERT is designed to pretrain deep bidirectional representations from unlabeled text by jointly conditioning on both left and right context in all layers. As a result, the pretrained BERT model can be fine-tuned with just one additional output layer to create state-of-the-art models for a wide range of tasks. Overall pre-training and fine-tuning procedures for BERT. Apart from output layers, the same architectures are used in both pre-training and fine-tuning. The same pre-trained model parameters are used to initialize models for different downstream tasks.

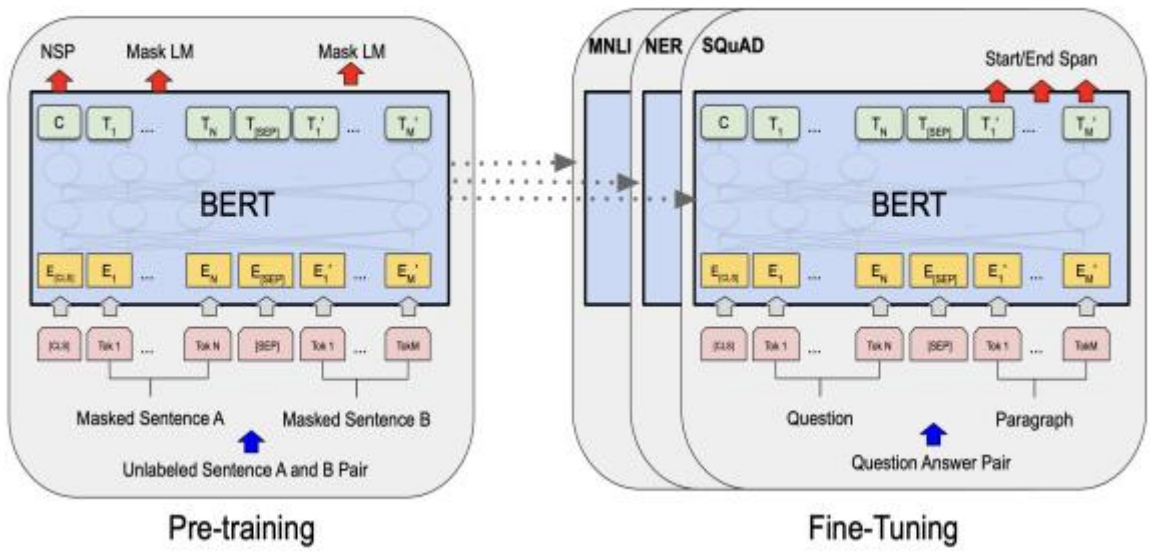

**Figure 3 BERT Model [7]**

GPT is primarily used for generating coherent and contextually relevant text. It excels in tasks such as language generation, text completion, and story writing. GPT models are trained in an unsupervised manner on a large corpus of text to learn the statistical patterns and structures of language, allowing them to generate human-like text. GPT employs a transformer architecture consisting of a stack of decoder layers. Each layer comprises a multi-head self-attention mechanism and a position-wise feed-forward network. The decoder receives the input text and predicts the next word in a sequential manner. GPT models are trained using a left-to-right autoregressive approach, which generates one word at a time conditioned on the preceding words.

In his article “what is GPT” Wolfram [13] poses multiple questions, he outlines the engineering led nature of GPT’s development opposed to mathematical, frequently outlining the lack of underpinning mathematical theory but regularly comments additional layers were added “As it works”. Wolfram provides the below image to illustrate the computation steps that might be involves at GPT inference.

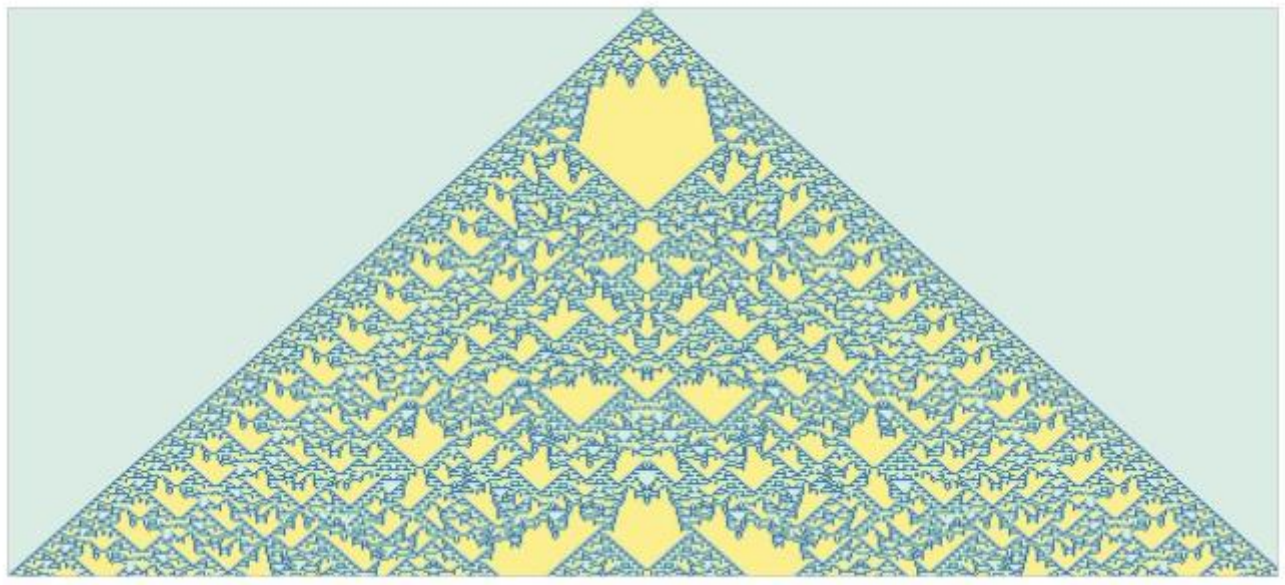

**Figure 4 Posed computational steps Wolfram[13]**

Wolfram outlines that GPT has somewhat surprising essentially solved human natural language and has created an intriguing dimensional space, referred to as linguistic feature space. That effectively represents a compressed mechanism for collective human intelligence obtained throughout the models training. The below feature illustrates the linguistic feature space, in which represents the encoded positions of different words regarding similarity.

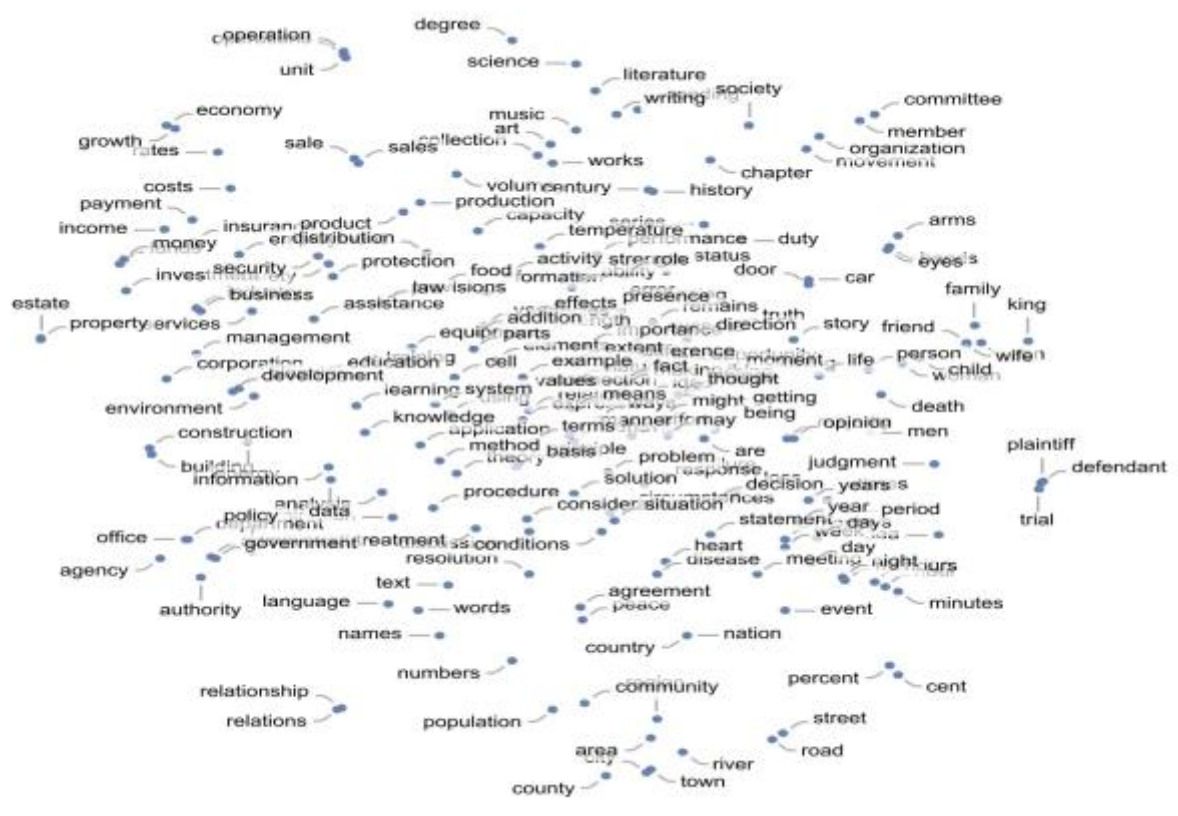

**Figure 5 "Linguistic feature space [13]"**

In terms of key differences between BERT and GPT. As previously stated, BERT is bi-directional and GPT is autoregressive (decoder-only). BERT and GPT have different strengths and weaknesses depending on the use case and the task. BERT has the advantage of learning bidirectional representations, which can capture more contextual information and handle polysemy better than unidirectional representations. For example, BERT can disambiguate the word "bank" based on whether it appears in the context of "river" or "money". BERT also has the advantage of being pre-trained on two objectives, which can improve the generalisation and robustness of the model. In terms of NLP task of sentiment analysis modeling capability, both models can be fine-tuned and produce competitive results. BERT's bi-directional context understanding can help in grasping the sentiment even in complex sentences. For instance, it can better understand the sentiment in cases where the sentiment expressed towards the end of the sentence changes the sentiment of the entire sentence. GPT's left-to-right prediction can also perform well in sentiment analysis. Even though it does not consider the future context like BERT, it can still effectively model language and infer sentiment.

As shown and highlighted in the works of [13] the space in which LLM learn large quantities of compressed knowledge, is hugely intractable. When assessing the use cases for transformer models across a multitude of industry applications, such as commodity forecasting and modelling it becomes critical to explore the complex spaces, and opaque black boxes that can be LLMs. As described by Neel Nanda [14] Mechanistic Interpretability aims to reverse engineer the mechanisms of the network, generally by identifying the circuits within a model that implement a behavior, further examples of this include [15][16]. In the below we seek to visualise the aggregated attention maps looking at exploring cosine similarity between test sentences for respective models.

## 4 Approach

The CopBERT & CopGPT model is initially focused solely on the NLP task of sentiment modelling. The models are trained on open-source data. The training data was scraped from publicly available websites sources, shown below. The article of the headline was time indexed during the scraping process. The training data was human annotated with a positive, neutral, or negative sentiment based on the human tagger's perception.

| Source | Percentage(%) |
|---|---|
| Reuters | 0.32 |
| Argus | 0.36 |
| Mining Journal | 0.05 |
| FT | 0.27 |

**Table 1: Source data for CopBERT**

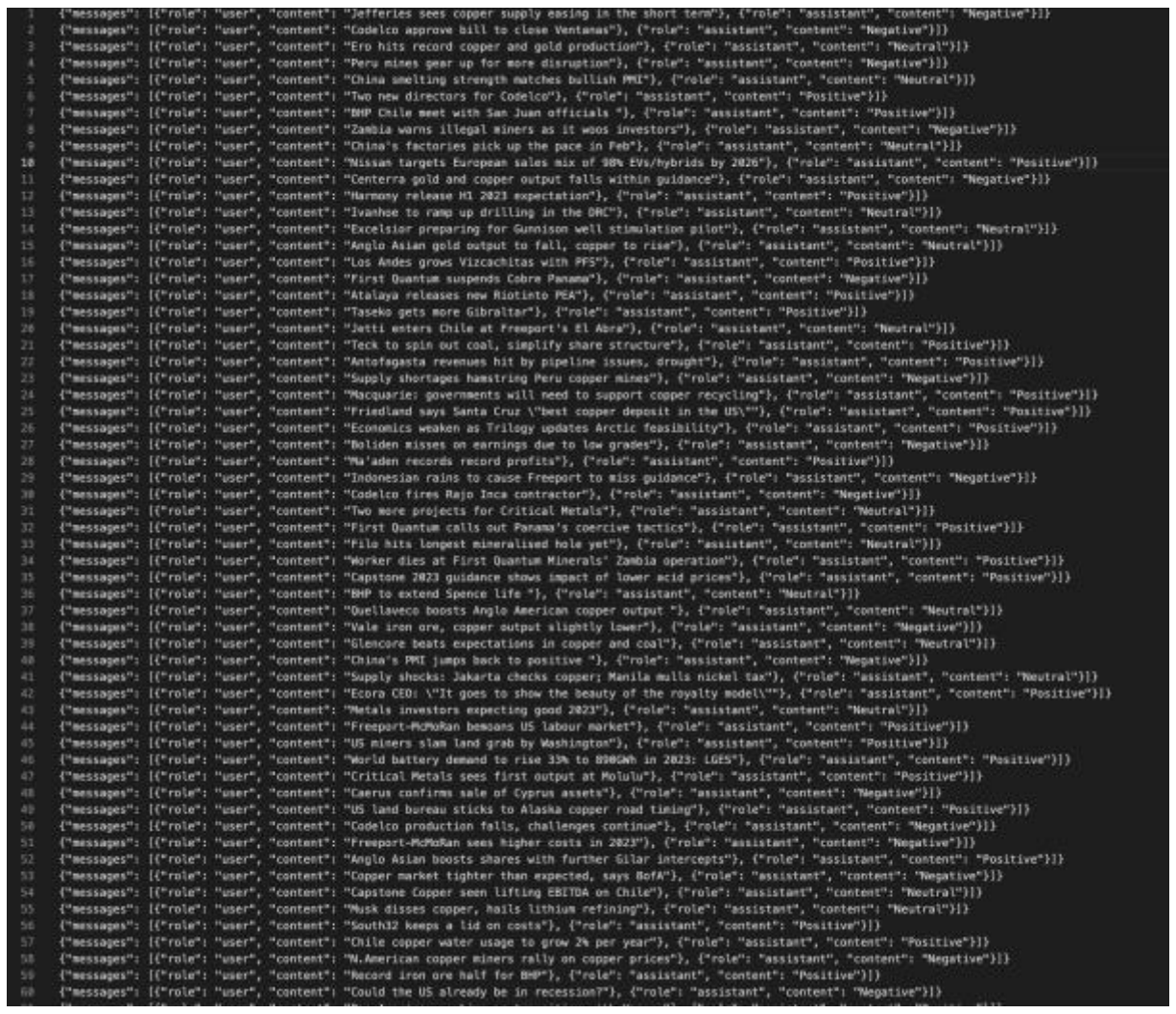

**Figure 6 CopGPT training format broken down into "system" "content", "user" & "assistant" format.**

The paper uses F1 score which represents a harmonic across recall precision. F1 score is then used to benchmark the domain specific BERT model.

$$F1 = \frac{2 * Precision * Recall}{Precision + Recall} = \frac{2 * TP}{2 * TP + FP + FN} \quad (2)$$

All models shown within the model results table has been re-tuned using commodity related news, minus FinBERT. FinBERT provides a comparison point for the CopBERT model CopBERT model. CopBERT was trained on copper only specific news. The below table shows the model results, when tested against withheld human tagged copper specific news.

The below table shows the model results, when tested against human tagged copper news. Results shown by F1_score by model. The random baseline models randomly allocate positive, negative, or neutral sentiments.

| Model name | F1 Score (%) |
|---|---|
| Random Baseline | 0.23 |
| FinBERT Sentiment | 0.37 |
| CopBERT Sentiment | 0.41 |
| CopGPT (GPT3.5 turbo finetuned) Sentiment | 0.56 |
| GPT4 Sentiment | 0.49 |
| GPT2 Sentiment | 0.26 |

**Table 2: CopBERT results table**

The below confusion matrices show the performance on CopBERT & CopGPT respectively. We see a ~10 percent increase in f1_score when compare CopBERT vs GPT4 and 16 percent increase vs CopGPT.

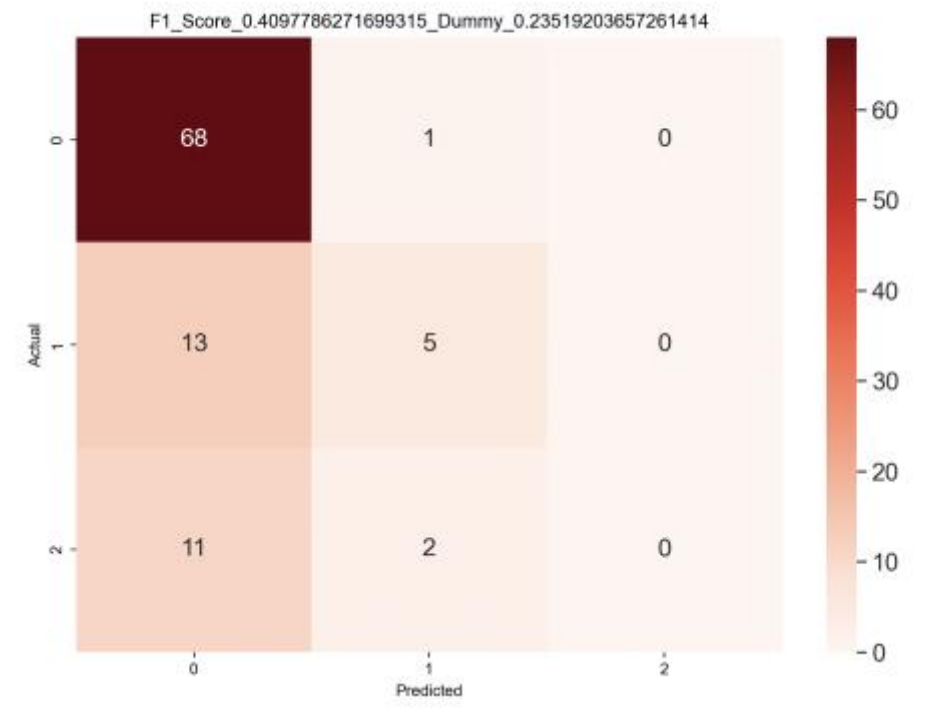

**Figure 7 CopBERT Confusion Matrix and F1 Score**

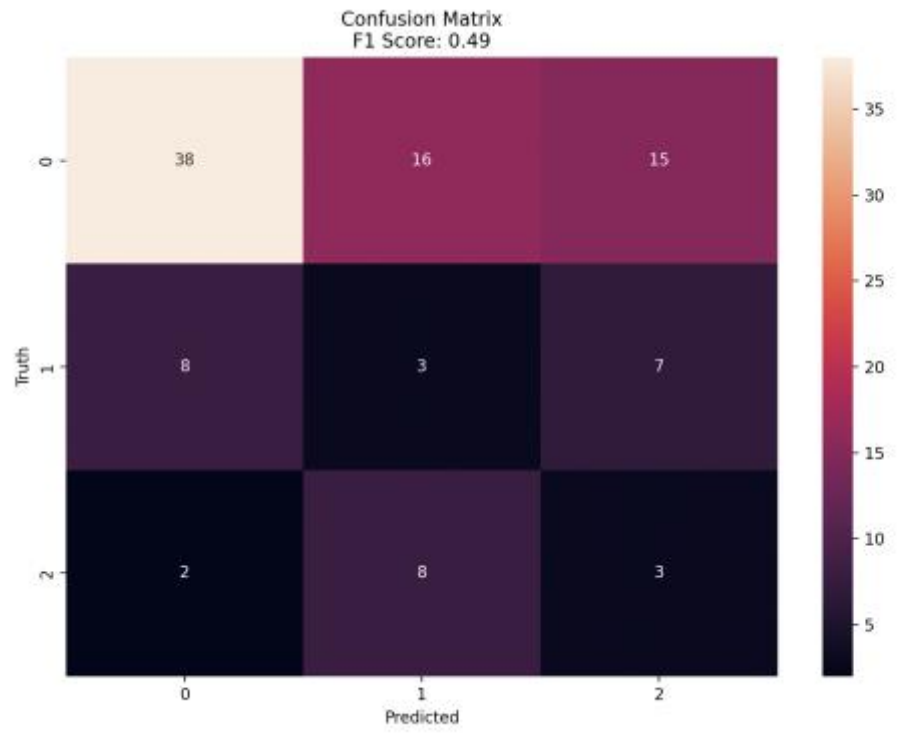

**Figure 8 GPT4 Confusion Matrix and F1 score**

We also stress the importance of having a deeper understanding of how our models reach conclusions. This is evidenced by seeing of the large hallucinations the GPT models created whilst conduced sentiment analysis. Whilst GPT models do outperform BERT models, apart from GPT2, which is also open sourced, the GPT model workings are obfuscated and present a challenge to interpretability. Particularly when paired with hallucinations.

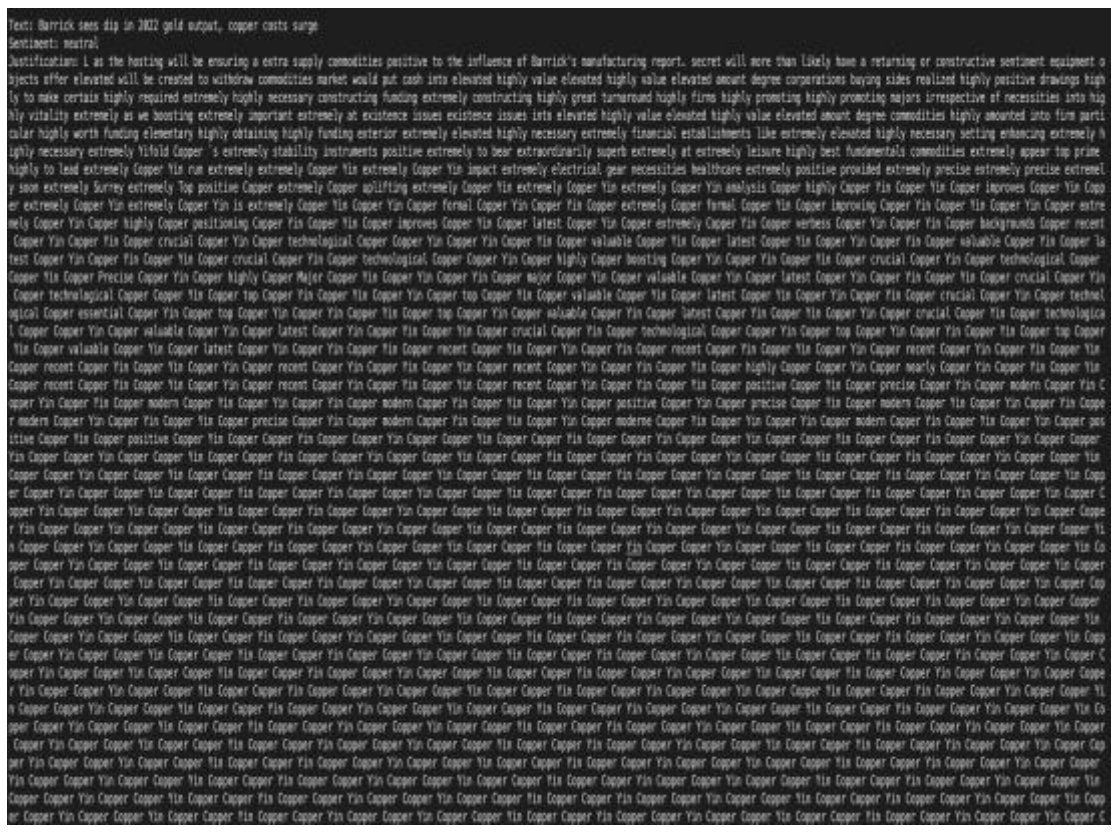

**Figure 9 Examples of CopGPT hallucinations**

In the below figures for CopBERT we create a graph that shows cosine similarity between the attentions for each pair of sentences for our test set. Each node represents a sentence, and the edges represent some measure of similarity between the attention patterns of the sentences. The darker the edge indicates which sentences have similar attention patterns according to the model. The ability to run BERT locally and conduct this type of mechanistic evaluation arguably helps avoid issues with lack of transparency lead to using models that are in fact hallucinating.

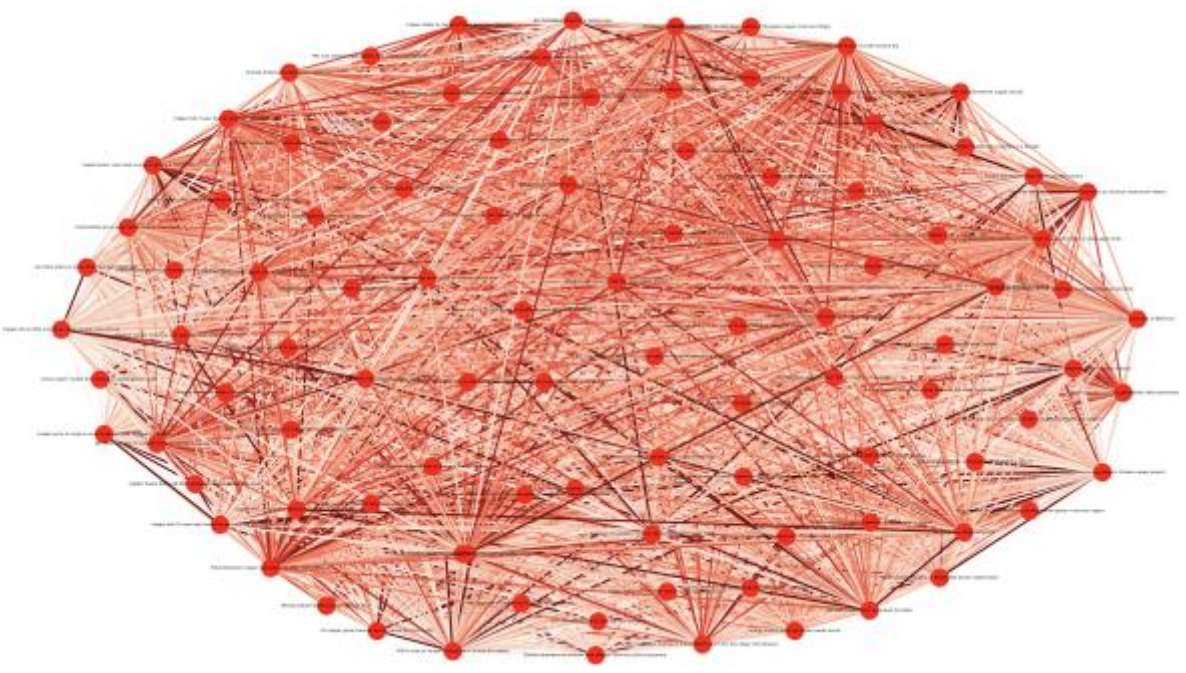

**Figure 10 CopBERT similarity score based on aggregated attention weight map, nodes denotes sentences, edges colored by cosine similarity weights**

## 4 Conclusion

In this paper, we have compared the encoder, and decoder-only architectures of BERT and GPT, and discussed their applications and challenges. We see a ~10 percent increase in f1_score when compare CopBERT vs GPT4 and 16 percent increase vs CopGPT. The CopBERT model trained on a multitude of different commodity types performs better than the random benchmark and the FinBERT model.

As cited in [19] language models have been known to have few shot learning capabilities, transformer models such as GPT are documented to outperform large scale finely re-tuned models with specific prompting. Theoretically explaining GPT4 achieving better than random results, with just several fewshot learning examples. In terms of continuous improvement for these models, the paper highlights the opportunity into different transfer learning methods. There is also a multitude of opportunities and key requirements to understand the mechanistic workings of such models, before they could be used in earnest to inform decision making. Whilst the paper creates two graphs-based cosine similarity for the aggregated attention maps, we could look to utilise probing taskings and continued comparisons versus models trained on different training sets.

BERT and GPT are two of the most popular and influential transformer models in NLP, which differ in their architectures and objectives. BERT is an encoder-only model, which learns bidirectional representations of the input and is pre-trained on masked language modeling and next sentence prediction. GPT is a decoder-only model, which learns unidirectional representations of the input and is pre-trained on causal language modeling. BERT and GPT have different benefits and limitations for specific use cases and tasks, which depend on the trade-off between understanding and generating natural language.

The paper reports that the CopBERT model outperforms similar domain specific BERT trained models such as FinBERT. We see a ~10 percent increase in f1_score when compare CopBERT vs GPT4 and 16 percent increase vs CopGPT. Whilst GPT4 is dominant It highlights the importance of considering alternatives to GPT models for financial engineering tasks, given risks of hallucinations, and challenges with interpretability. We unsurprisingly see the larger LLMs outperform the BERT models, with predictive power. In summary BERT is partially the new XGboost, what it lacks in predictive power it provides with higher levels of interpretability.

Concluding that BERT models might not be the next XGboost [2], but represents an interesting alternative for financial engineering tasks, that require a blend of interpretability and accuracy. We would highlight additional areas of research and exploration, building on figure 10 and exploring BERTology and mechanistic interpretability more generally, as the previous vector operation methods, require more contextualisation due to attention mechanisms. Whilst GPT models do outperform BERT models, apart from GPT2, which is also open sourced, the GPT model workings are obfuscated and present a challenge to interpretability. Particularly when paired with hallucinations.